\documentclass[prl,superscriptaddress,twocolumn]{revtex4}
\usepackage{epsfig}
\usepackage{graphics}
\usepackage{graphicx}

\begin{document}

\title{Strong diamagnetic response and specific heat anomaly above $T_c$ in underdoped $\mbox{La}_{2-x}\mbox{Sr}_x\mbox{CuO}_4$}

\author{U. Thisted}
\affiliation{Department of physics, Norwegian University of Science and Technology, N-7491 Trondheim, Norway}
\author{J. Nyhus}
\affiliation{Department of physics, Norwegian University of Science and Technology, N-7491 Trondheim, Norway}
\author{T. Suzuki}
\affiliation{Department of Quantum Matter, ADSM, Hiroshima University, 
1-3-1 Kagamiyama, Higashi-Hiroshima 739-8526, Japan}
\author{J. Hori}
\affiliation{Department of Quantum Matter, ADSM, Hiroshima University, 
1-3-1 Kagamiyama, Higashi-Hiroshima 739-8526, Japan}

\author{K. Fossheim}
\affiliation{Department of physics, Norwegian University of Science and Technology, N-7491 Trondheim, Norway}

\begin{abstract}

By measuring AC susceptibility using a very low amplitude of the AC field ($<1$ mG) it is shown that  underdoped samples of $\mbox{La}_{2-x}\mbox{Sr}_x\mbox{CuO}_4$, are diamagnetic in a temperature region above $T_c$ up to a temperature $T^*$. This behavior is only observed with AC fields along the c-axis whereas for fields in the ab-plane no diamagnetism above $T_c$ was detected. The diamagnetism is almost frequency independent in the frequency range 0.1-10 kHz.
At $T^*$ a broad step anomaly in the specific heat is inferred through measurements of the elastic constant $c_{33}$. We suggest that the observed diamagnetism and the anomaly in the elastic constant are associated with the existence of phase incoherent Cooper pairs between $T_c$ and $T^*$.
\end{abstract}

\maketitle


\section{introduction}
The nature of the so-called $T^*$-line in the high-$T_c$ cuprate superconductor phase diagram for $T$ vs. doping is among the most intensely debated issues in high-$T_c$ superconductor research in recent years. A pseudogap in the electronic excitation spectrum has been observed below $T^*$ by many different techniques \cite{timusk}. For most high-$T_c$ superconductors the pseudogap is strongest in underdoped samples and rapidly decreases with doping before terminating somewhere in the overdoped region \cite{fischer}. For $\mbox{La}_{2-x}\mbox{Sr}_x\mbox{CuO}_4$ however, a pseudogap has only been found in the underdoped region \cite{naeini}. The connection between the superconducting gap and the pseudogap has been a point of discussion. Several groups have obtained result indicating that Cooper pairs exist between $T^*$ and $T_c$, as a precursor to the Meissner phase. Recently Iguchi et.al. found diamagnetic regions above $T_c$ in thin film and single crystal of underdoped LSCO using a scanning SQUID microscope \cite{iguchi}. Also, vortex-like excitations above $T_c$ in (LSCO) have been reported by Xu et. al. \cite{xu}. Here we report on magnetic susceptibility measurements showing a strong diamagnetic response. In addition, we report an anomaly in the elastic constant $c_{33}$ at the temperature where we observe the onset of the diamagnetic response. This anomaly is thermodynamically related to a specific heat anomaly, on general grounds which will be discussed below. These findings have interesting implications for the understanding of underdoped high-$T_c$ superconductors in the temperature range between $T_c$ and $T^*$.


\section{experimental}
Three samples of $\mbox{La}_{2-x}\mbox{Sr}_x\mbox{CuO}_4$, all grown by the TSGZ method \cite{takashi}, have been investigated. All three samples have a Strontium content close to the optimal value of $x=0.15$. As measured by EPMA the values for the three samples are 0.146, 0.145 and $0.146\pm 0.002$ assigned as sample 1, 2 and 3 respectively. The sample sizes were (2x2x2)mm for sample 1 and 2, and (1.5x1.5x1)mm for sample 3.

The AC susceptibility ($\chi_{{}_{AC}}$) measurements were performed using a PPMS from Quantum Design, which was modified in order to be able to measure at lower AC amplitudes, currently down to 0.18 mG. The temperature was stabilized at each measurement, and temperature fluctuations are less than $\pm 0.03$ K, which was sufficient for the present purpose. The 5-point measurement technique of the PPMS reduces background contributions to a minimum.

Magnetisation measurements were performed using the moving sample technique of the PPMS. For this type of measurements the temperature variations were less than $\pm 5$ mK.

The elastic stiffness data were obtained using a high resolution ($10^{-6}$) ultrasonic continuous wave method \cite{bolef}\cite{jorgen} operating at a carrier frequency of 95 MHz. In this case $\sim$1 mK temperature stabilization was obtained.


\section{results}
\begin{figure}[htb]
\begin{center}
\includegraphics[width=8.5cm]{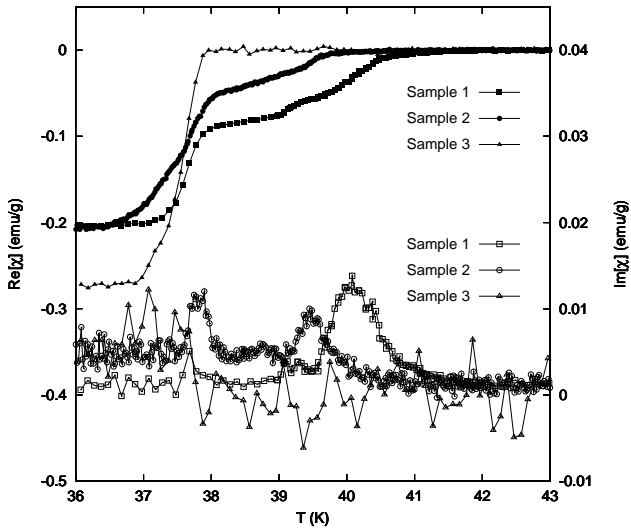}
\end{center}
\caption{\label{sammenlign} Real (filled symbols) and imaginary (open symbols) part of AC-susceptibility for samples 1-3. Measurements are performed at f=10kHz and with an AC field amplitude of 0.18 mG applied along the c-axis. Lines are guides to the eye.}
\end{figure}

\begin{figure}[htb]
\begin{center}
\includegraphics[width=8.5cm]{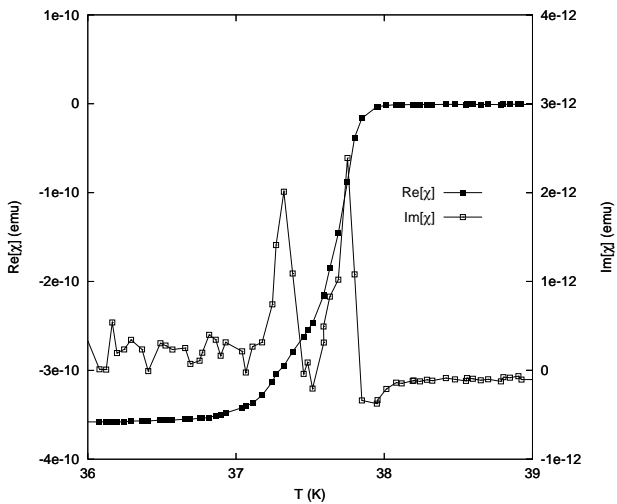}

\end{center}
\caption{\label{abretning} Real (filled symbols) and imaginary (open symbols) part of AC-susceptibility for sample 1 with the field applied perpendicular to the c-axis. Lines are guides to the eye. Notice the two distinct peaks in the imaginary part clearly indicating two domains in the sample. The calibration of this measurement was different than the other measurements.}
\end{figure}

$\chi_{{}_{AC}}$ of each of the three samples was measured in a frequency range of 0.1-10 kHz and at AC field amplitudes of $0.18$ mG to 1 G with the field aligned along the c-axis. Results at 10 kHz and 0.18 mG are displayed in Fig. \ref{sammenlign}. Sample 3, which we suggest is slightly overdoped, shows the behavior typical for high amplitude (1-10 G) $\chi_{{}_{AC}}$ measurements. Sample 1 and 2, however, show a different behavior. At a temperature $T^*$ well above $T_c$, it gives a strong diamagnetic response, before becoming perfectly diamagnetic at $T_c$. For sample 1 we see two steps above $T_c$ indicating that the sample consists of two domains with slightly different doping, this fact is also supported by the data in Fig. \ref{abretning}. The imaginary part of $\chi_{{}_{AC}}$ shows distinct peaks at $T^*$, these are located at 39.2 K and 40.1 K for the two domains of sample 1 and at 39.5 K for sample 2, and match the steps in the real part of $\chi_{{}_{AC}}$. We suggest that this different behavior is due to slightly different doping levels with samples 1 and 2 being underdoped. For all three samples $T_c$ is between 37 and 38 K, indicating that their doping content is close to optimal doping.

\begin{figure}[htb]
\begin{center}
\includegraphics[width=8.5cm]{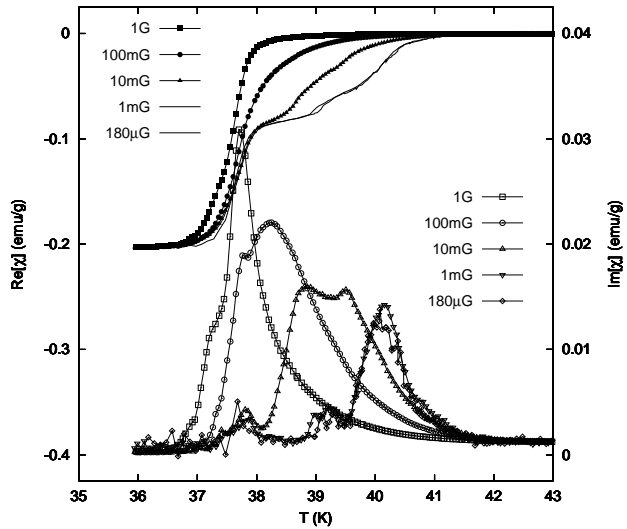}
\end{center}
\caption{\label{amplitude} Real (filled symbols) and imaginary (open symbols) part of AC-susceptibility for sample 1 when measured with different AC amplitudes. Measurements are performed at f=10kHz and with the field applied along the c-axis. The 1 mG and 180 $\mu$G curves almost coincide, and are shown without data markers for clarity. The 180 $\mu$G curve is slightly lower at 39 K. Lines are guides to the eye.}
\end{figure}

To demonstrate that the diamagnetic response above $T_c$ is strongly dependent on the AC field amplitude we show in Fig. \ref{amplitude} data for $\chi_{{}_{AC}}$ of sample 1 measured with different amplitudes of the AC field. First, let us point out that the temperature at which a perfectly diamagnetic response is obtained rises slightly with decreasing amplitude. 
The "onset" temperature of the diamagnetic response increases more rapidly, and for the lowest field measurements $Re\left[\chi_{{}_{AC}}\right]$ starts to develope a plateau above 38 K, with two distinct steps above $T_c$ due to the two domains of the sample. The peak in the imaginary part of $\chi_{{}_{AC}}$ at 37.7 K remains constant in temperature but reduces significantly in height with decreasing amplitude. Furthermore, another peak in $Im\left[\chi_{{}_{AC}}\right]$ separates from it and shifts towards higher temperatures as the amplitude is decreased. For all measurements, the peaks in the imaginary part are located at the same temperatures as the steps in the real part. 
The data obtained with 1 mG and 180 $\mu$G amplitudes (no markers on $Re\left[\chi_{{}_{AC}}\right]$ in Fig. \ref{amplitude}) are practically indistinguishable, and imply that there is an intrinsic limiting low-amplitude behavior.

In contrast, if the AC field is directed perpendicular to the c-axis, we observe no diamagnetic response above $T_c$, as shown in Fig. (\ref{abretning}). Here, we again see a double transition at $T_c$ indicating that sample 1 consists of two domains of slightly different doping.

\begin{figure}[htb]
\begin{center}
\includegraphics[width=8.5cm]{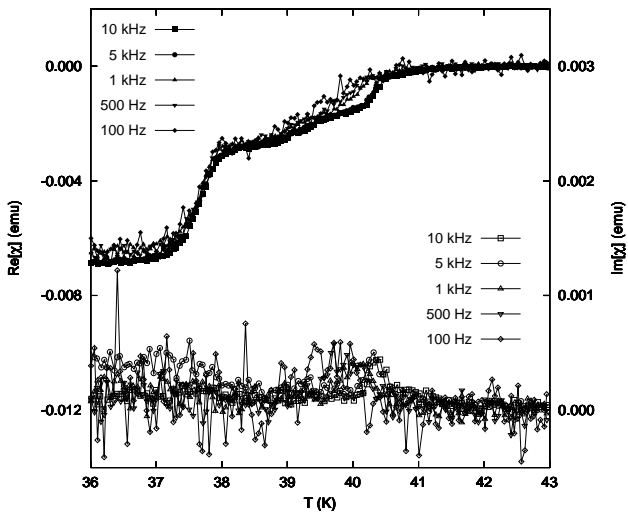}
\end{center}
\caption{\label{frekvens} Real (filled symbols) and imaginary (open symbols) part of AC-susceptibility for sample 1 when measured at different AC frequencies. Measurements are performed with an amplitude of 180 $\mu$G and with the field applied along the c-axis. The curves show a very small dependence of $\chi_{{}_{AC}}$ on frequency, and only for temperatures just below $T^*$ is the difference distinctly higher than the noise.}
\end{figure}

$\chi_{{}_{AC}}$ in the $\vec{H}\parallel\vec{c}$ configuration was found to be almost frequency independent in the range 0.1-10 kHz as shown in Fig. (\ref{frekvens}). There is only a small difference for temperatures slightly below $T^*$ for which the there is an increasing diamagnetic response with increasing frequency. However, the curves merge at the same level of the plateau between 38 K and 39 K. The frequency independence in this range shows that the screening currents have very long or infinite decay time.
Due to increasing signal to noise ratio with increasing frequency, 10 kHz was chosen for all other measurements displayed in the figures. 

The results in Fig. (\ref{sammenlign}),  Fig. (\ref{frekvens}) and Fig. (\ref{abretning}) indicate that in the temperature range between $T_c$ and $T^*$ it is possible to induce Meissner-like screening currents in the Cu-O planes, whereas this is not possible in the direction across the Cu-O planes. 

\begin{figure}[htbp]
\begin{center}
\includegraphics[width=8.5cm]{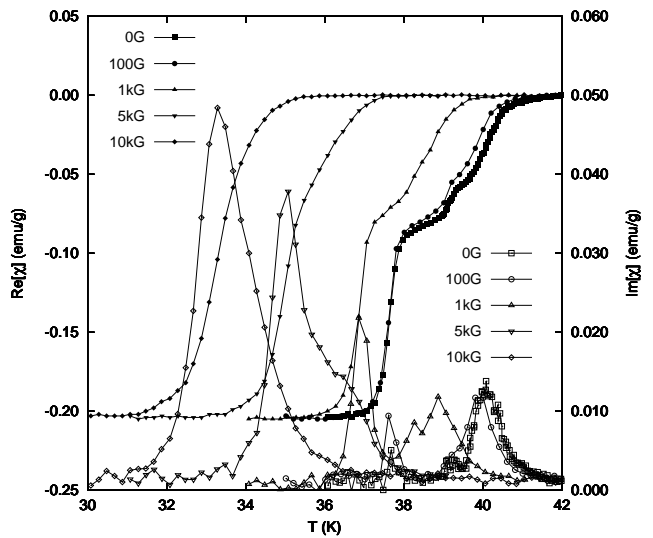}
\end{center}
\caption{\label{bakgrunn} Real (filled symbols) and imaginary (open symbols) part of AC-susceptibility for sample 1 in different DC bias fields. Measurements are taken during field cooled conditions at f=10kHz and with an amplitude of 0.18 mG. Both AC and DC fields are applied along the c-axis. Lines are guide to the eye.}
\end{figure}

In addition, we have investigated how a DC magnetic field influences $\chi_{{}_{AC}}$. The results displayed in Fig. \ref{bakgrunn} show that even though an AC field of 1 G completely suppresses the diamagnetism above $T_c$, a parallel DC field of 100 G only gives a small shift towards lower temperature of the transition at $T^*$. We see that when increasing the DC field, $T^*$ decreases faster than $T_{irr}$ ($T_{irr}$ being the temperature where $\chi_{{}_{AC}}$ reaches its value of perfect diamagnetism), and that at 10 kG the two have almost merged. We also note that the peak height at $T_{irr}$ in the imaginary part of $\chi_{{}_{AC}}$ increases when the field is increased.

When analysing large amplitude $\chi_{{}_{AC}}$ data, it is usually assumed that the response is near zero above the irreversibility line (IL), do to the free flow of vortex lines. In the vicinity of the IL we observe the onset of the diamagnetic response, and a peak in the imaginary part of $\chi_{{}_{AC}}$ caused by irreversible vortex motion. Far below the IL there is no (or very little) flow of vortices in and out of the sample, and hence the response is perfectly diamagnetic with zero imaginary part. 

The diamagnetic response in the low amplitude $\chi_{{}_{AC}}$ measurement between $T_c$ and $T^*$ indicate that vortices do not flow freely into the sample, and we suggest that there is a barrier to vortex penetration. The Bean-Livingston surface barrier \cite{bean-livingston} or the geometrical barriers discussed by Brandt \cite{brandt} and Zeldov et.al. \cite{zeldov} presents such barriers to penetration. The peak in the imaginary part of $\chi_{{}_{AC}}$ for low amplitude measurements at $T^*$ is then caused by hysteresis due to this barrier. 

It has been shown that the line tension of vortices vanish above a field dependent temperature, that in the zero field case is $T_c$\cite{kiet}. This implies that point defects (such as Strontium sites) will not act as pinning centers. Therefore, surface barriers seem to be the only possible explanation for limitations to the vortex movement.

A very small increase in the magnetic field would induce the Cooper pairs to set up screening currents. If the magnetic field increase is not sufficient to overcome the barrier and change the number of vortices in the sample, the response will be diamagnetic. However, if the magnetic field increase is larger, vortices will be pushed into the sample, and no global diamagnetic response will be observed. This can explain the presence of a small frequency dependence in the temperature region just below $T^*$. The surface barrier is lower at these temperatures due to lower cooper pair density, and thermally induced jumping of vortices over the barrier is facilitated. This will lead to a weaker diamagnetic response for lower frequencies.

The apparent position of $T^*$ is then the temperature where the amplitude of the AC field becomes smaller than the field required to overcome the barrier. The apparent decrease in $T^*$ with increasing fields is then possibly, at least in part, caused by a reduction of the vortex penetration barrier as shown by Bean and Livingston \cite{bean-livingston}.  
Actually, it is the strong enhancement of the diamagnetic response below $T_c$ which is most surprising, we point to several possible explanations for this: i)A reduction of the Cooper pair density leading to weaker screening currents. ii)The presence of a thermally induced vortex tangle \cite{kiet} leading to weaker screening, because the currents associated with the vortices reduce the possible screening currents. iii)Only parts of the sample behaves diamagnetic above $T_c$ as indicated by the results obtained by Iguchi et.al.\cite{iguchi}.

\begin{figure}[hbtp]
\begin{center}
\includegraphics[width=8.5cm]{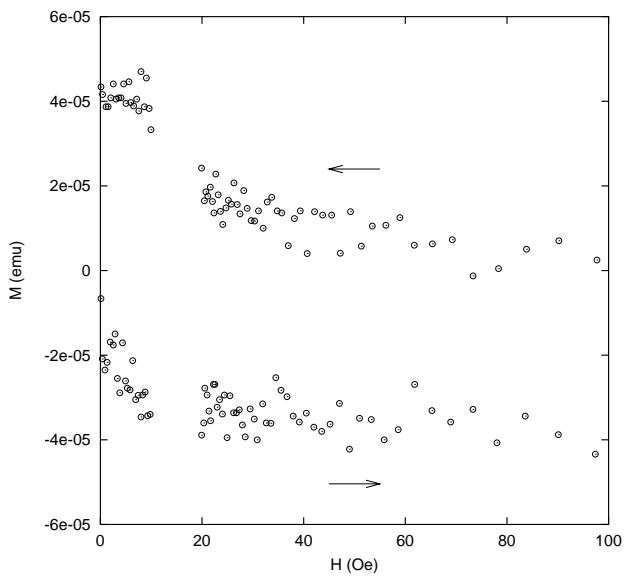}
\end{center}
\caption{\label{DC} Magnetisation measurements of sample 1, at constant temperature $T=38$ K with field along the c-axis. A background which is linear in $H$ and is due to the sampleholder has been subtracted. Arrows indicate the directions of the field sweep. The field was raised to 10 kOe between measurement of increasing and decreasing branches.}
\end{figure}

Measurements of the DC magnetisation were also done to probe the static behavior. Results obtained at 38 K for sample 1 are presented in Fig. \ref{DC}. We find a clear hysteretic behavior, although the magnetic response is very weak. A background linear in $H$ arising from the sampleholder has been subtracted. This background shows no hysteresis, neigther does a similar measurement of the sample performed at 50 K. These result agree well with the AC results, and also further support our suggestion of a surface barrier.

\begin{figure}[htb]
\begin{center}
\includegraphics[width=8.5cm]{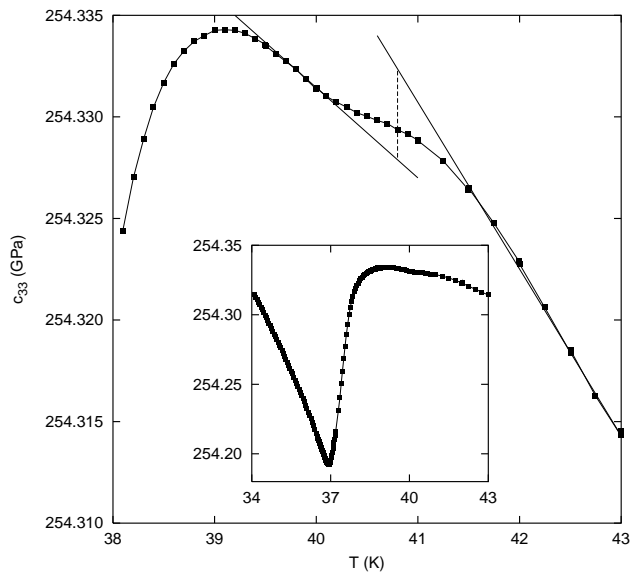}
\end{center}
\caption{\label{c33} Elastic constant $c_{33}$ for sample 1 measured by ultrasound. There is a small but significant step feature at $T^*$ between 40 K and 41 K. The large cusp anomaly at $T_c$ is shown in the insert. Lines are guides to the eye.}
\end{figure}

From basic thermodynamic relations it follows, that under certain conditions, an anomaly in the specific heat at a phase transition results in a corresponding anomaly in the elastic constants \cite{testardi}. This has been experimentally verified for the fluctuation peak in $C_p$ at $T_c$ for $\mbox{La}_{2-x}\mbox{Sr}_x\mbox{CuO}_4$, sample 1 \cite{thisted}. As can be seen in Fig. \ref{c33} there is an anomaly in $c_{33}$ at $T^*$. 
The sharp minimum in $c_{33}$, seen in the inset of Fig. \ref{c33}, is caused by critical fluctuations, which are transverse phase-fluctuations of the superconducting order parameter $\psi = \left|\psi\right| e^{i\Phi}$\cite{kiet}\cite{tesanovic}\cite{kivelson}. The transition at $T^*$ is broad and smooth, consistent with non critical gaussian fluctuations of the amplitude of the order parameter.

In our view the observed doubble transition cannot be caused by impurities or multiple domains in the sample. The highest reported $T_c$ of $\mbox{La}_{2-x}\mbox{Sr}_x\mbox{CuO}_4$
regardless of doping is 38 K, the transition at $T^*$ in sample 1 is above 40 K, and for sample 2 just below 40 K. Furthermore, if the two transitions were caused by different domains we would expect to observe the following in the temperature region between 38 K and 40 K: i) A diamagnetic response for $\vec{H}\perp\vec{c}$ not only for $\vec{H}\parallel\vec{c}$. ii) Only a weak amplitude dependence $\chi_{{}_{AC}}$. iii) A similar behaviour of the peak height at the two transitions with increasing background field. Since none of these effect are observed we conclude that the two transitions arise from different physical mechanism.


\section{conclusion}
We have demonstrated that, by performing AC susceptibility with very low AC field amplitudes, it is possible to measure a diamagnetic response of weakly underdoped high-$T_c$ superconductors, above $T_c$. The diamagnetic response which has only been detected for fields along the c-axis exist up til a temperature $T^*$ above which it drops to zero. In addition at $T^*$ there is a step in the specific heat, as found through measurement of $c_{33}$. 

Our results show a clear difference between the two transition at $T_c$ and $T^*$. The changes in $\chi_{{}_{AC}}$ at $T^*$ are sensitive to changes in the AC amplitude, while at $T_c$ they are almost independent of the AC field amplitude. At $T^*$ there is a large peak in the loss component of $\chi_{{}_{AC}}$, whereas this is almost absent at $T_c$. We found that $\chi_{{}_{AC}}$ is frequency indepent except for temperatures just below $T^*$. This is also supported by the observation of magnetic hysteresis above $T_c$ in DC magnetisation. Furthermore, the $c_{33}$ data show a peak due to critical fluctuations at $T_c$ distinctly different from the step at $T^*$.

We have also shown that the diamagnetism above $T_c$ exists in the presence of a strong background field. We suggest that our data can be explained by the presence of phase incoherent Cooper pairs between $T_c$ and $T^*$, and that there is a barrier to vortex penetration causing the difference between high and low amplitude $\chi_{{}_{AC}}$ measurements.


\begin{acknowledgments}
The authors would like to thank A. Sudb{\o} for valuable discussions. This work is supported by the Research Council of Norway and The ESF Vortex programme.
\end{acknowledgments}

\end{document}